\documentclass[twocolumn,pre,aps]{revtex4-1}
\usepackage[T1]{fontenc}
\usepackage[latin9]{inputenc}
\usepackage{graphicx}
\usepackage[unicode=true,pdfusetitle,
 bookmarks=true,bookmarksnumbered=false,bookmarksopen=false,
 breaklinks=false,backref=false,colorlinks,
 citecolor=blue, linkcolor=blue]
 {hyperref}

\usepackage{rotating}

\makeatletter
\usepackage{amssymb}
\usepackage{todonotes}

\usepackage{multirow}
\makeatother

\begin{document}

\title{Scaling laws in critical random Boolean networks with general in- and out-degree distributions}

\author{Marco M\"oller and Barbara Drossel}

\affiliation{Institute for condensed matter physics, TU Darmstadt, Hochschulstra\ss e 6, 64289 Darmstadt, Germany }

\begin{abstract}

We evaluate analytically and numerically the size of the frozen core and various scaling laws for critical Boolean networks that have a power-law in- and/or out-degree distribution. To this purpose, we generalize an efficient method that has previously been used for conventional random Boolean networks and for networks with power-law in-degree distributions. With this generalization, we can also deal with power-law out-degree distributions. When the power-law exponent is between 2~and~3, the second moment of the distribution diverges with network size, and the scaling exponent of the nonfrozen nodes depends on the degree distribution exponent. Furthermore, the exponent depends also on the dependence of the cutoff of the degree distribution on the system size. Altogether, we obtain an impressive number of different scaling laws depending on the type of cutoff as well as on the exponents of the in- and out-degree distributions. We confirm our scaling arguments and analytical considerations by numerical 
investigations.

\end{abstract}

\maketitle

\section{Introduction}

Boolean networks are often used as generic models for the dynamics of
complex systems such as social and economic networks, neural networks,
and gene or protein interaction networks
\cite{kauffman:random,kauffman:metabolic}. Whenever the states of the
constituents of the system can be reduced to being either ``on'' or
``off'' without loss of important information, a Boolean approximation
captures many features of the dynamics of real networks
\cite{BornholdtScience}. In order to understand the generic behavior
of Boolean networks, random models were investigated in depth
\cite{reviewbarbara}, although it is clear that neither the connection
pattern, nor the usage of update functions of biological networks is
reflected realistically in such random models. For random models,
the distinction between frozen, chaotic, and critical networks has
become commonplace \cite{kauffman:homeostasis,Kauffman1969437}, with
frozen networks having short attractors where almost all nodes are fixed at one value, while chaotic networks have attractors the length of which increases exponentially with the system size. Critical networks, which are ``at the edge of chaos'', are considered particularly relevant for biological systems, and for this reason many investigations have concentrated on such critical networks. One important result of these investigations was that the number of nodes that do not become frozen on the attractors increases as $N^{2/3}$ with the network size $N$ \cite{socolar:scaling,kaufmanandco:scaling,TamaraContainerAnalytisch}.

The majority of recent research on Boolean networks was devoted to
networks with more realistic features. In particular, many natural
networks are scale free, which means that the number of connections
per node follows a power law \cite{albert2005scale}. Typically, the
power-law exponent of the degree distribution is between 2 and 3,
which means that the second moment of this distribution diverges with
network size. Due to their relevance to natural systems, scale-free
Boolean networks have been investigated by various authors. In order
to keep these models as simple as possible, connections between nodes
are made at random within the constraints given by the degree
distribution. The Boolean functions are usually also assigned at
random from a certain set of functions, e,g., threshold functions.
Observations made in computer simulations for these networks are that
attractors are shorter and frozen nodes are more numerous in critical
scale-free networks compared to conventional random Boolean networks
with the same total number of links and of nodes
\cite{fox.hill:from,kinoshita.iguchi.ea:robustness}, that attractors
are sensitive to perturbations of highly connected nodes, but not of
sparsely connected nodes
\cite{aldana03,kinoshita.iguchi.ea:robustness}, and that scale-free
Boolean networks evolve much faster and more steadily towards a target
pattern of an ``output'' node than conventional random Boolean
networks \cite{oikonomou2006effects}.  Analytical results are mostly
limited to calculating the phase diagram using the annealed
approximation
\cite{aldana03,2003PNAS..100.8710A,fronczak.fronczak.ea:kauffman}. 
Similarly to conventional random Boolean networks, which have a fixed in-degree, scale-free Boolean networks show also frozen, critical, or chaotic dynamics, depending on the parameter values.
 A
comparison between results obtained from the annealed approximation
and from computer simulations is performed in
\cite{iguchi2005boolean}, giving not always an agreement between the
two approaches. Three years ago, Drossel and Greil
\cite{drossel2009critical} derived analytically the scaling exponents
for the number of nonfrozen nodes in critical Boolean networks with a
scale-free in-degree distribution. The values of these exponents
depend continuously on the exponent of the in-degree distribution, if
it is smaller than 3. So far, the equivalent case of scale-free
out-degree distributions was not yet investigated, although it is
relevant for natural systems \cite{albert2005scale}. From computer
simulations of Boolean networks with a scale-free out-degree
distribution, it is known that the properties of attractors are
different from the case of a scale-free in-degree distribution
\cite{2004PhyA..339..665S}.

It is the purpose of this paper to derive the scaling exponents of the number of nonfrozen nodes for Boolean networks that have a power-law out-degree distribution. To this goal, we generalize an efficient method that has previously been used for conventional random Boolean networks \cite{kaufmanandco:scaling,TamaraContainerAnalytisch} and for networks with power-law in-degree distributions \cite{drossel2009critical}. The result is a stunning variety of scaling laws depending on the in- and out-degree exponent as well as on the type of cutoff used in both cases. In particular, we also find that the dependence of the scaling exponents on the degree distribution shows opposite trends  for scale-free in-degree and for scale-free out-degree distributions. For one of the many cases investigated by us, we obtain the scaling law given in \cite{lee2008broad}. However, we find this scaling law for a different situation than the one considered in \cite{lee2008broad}.
We confirm our analytical results by computer simulations.

\section{Model}

We consider Boolean networks consisting of $N$ nodes, where each node~$i$ has a Boolean value~$\sigma_i \in \{0,1\}$ and is connected to other nodes via its in- and outputs. Furthermore, each node is assigned a Boolean update function.  Connections and update functions are chosen at random, given the distribution $P(k_\mathrm{in})$ and $P(k_\mathrm{out})$ of the number of inputs and outputs of the nodes, and the distribution of update function. 

In conventional random Boolean networks, all nodes have the same number $k$ of inputs, $P(k_\mathrm{in})= \delta_{k_\mathrm{in},k}$ and a Poisson distribution of the number of outputs, 
\begin{equation}
  P(k_\mathrm{out})= \frac {k^{k_\mathrm{out}}}{k_\mathrm{out}!}e^{-k}\, , 
\end{equation}
since each node receives its input from each other node with the same probability. Different probability distribution for Boolean functions are used, with \emph{biased functions} being a familiar choice, where for each possible combination of the $k_i$ input values the output is 1 with probability $p$ and 0 with probability $1-p$. By adjusting the value of $p$, the network can be made critical. In this paper, we use only \emph{constant} and \emph{reversible} functions. For each value of the in-degree, there are two constant functions, which take the same value for all possible inputs, and two reversible functions, which are defined by the condition that the change of one input always changes the output. For $k_\mathrm{in}=2$, these functions are XOR and NOT XOR. We will argue in 
the end that our results are also valid for other choices of update functions, in particular for biased functions.

In this paper, we consider scale-free networks, where either the distribution of inputs is a power law, $P(k_\mathrm{in}) \propto k_\mathrm{in}^{-\gamma_\mathrm{in}}$ or the distribution of outputs is a power law, $P(k_\mathrm{out}) \propto k_\mathrm{out}^{-\gamma_\mathrm{out}}$, or both are a power law. We only consider the case that there is no correlation between the  in-degree and the out-degree of a node. We focus on the interesting case of $\gamma_\mathrm{in},\gamma_\mathrm{out}  \in \left(2,3\right)$, where the second moment of  $k_\mathrm{in}$ or $k_\mathrm{out}$ diverges, but the mean value is well defined. The mean values of the in- and out-degree distribution have to be identical, since the total number of inputs and outputs must be the same because each link connects an input with an output. 

We generated scale-free in- or out-degree distributions $P(k) \propto k^{-\gamma}$ in two ways, which lead to a different scaling of the  cutoff values $k^\mathrm{max}(N)$ with network size $N$. One the one hand, we draw each value $k_i$ at random from the distribution $P\left(k\right)$, which leads to a cutoff 
\begin{equation}
k^\mathrm{max}\propto N\, . \label{kmax1}
\end{equation} 
Alternatively, we fixed the number of nodes with the in- or output value $k$ to exactly $c\cdot N \cdot P\left(k\right)$ rounded to the next integer, while adjusting $c\gtrapprox 1$ such that the size of the sample is as close as possible to $N$. This leads to a cutoff 
\begin{equation}
k^\mathrm{max}\propto N^{\frac{1}{\gamma}}\, . \label{kmax2}
\end{equation} 
We fixed the minimum value of $k$ in scale-free distributions to 2, since this leads to  more nodes with larger values of $k$ and therefore to a faster approach to the asymptotic behavior with increasing $N$. Furthermore, we adjusted the number of nodes with $k=2$ connections such that the mean value of $k$ does not change with $N$, but equals the asymptotic value taken in the limit $N \to \infty$. This also reduces finite-size effects.

In order to make the networks critical, the proportion $r_\mathrm{c}$ of reversible functions was chosen such that the change of the state of one node propagates on average to one other node, implying
$  \left(1-r_\mathrm{c}\right)\cdot 0 + r_\mathrm{c}\cdot \left< k \right> = 1$, and leading to
\begin{equation}
  r_\mathrm{c} = \frac{1}{\left< k \right>} .\label{criticality}
\end{equation}
For such critical networks, the number of nodes that do not become frozen when the system is on an attractor increases sublinearly with network size, with a power-law exponent that will be determined below.

\section{Algorithm: Determining the frozen core starting from constant functions}

An elegant way to determine the frozen core was suggested in
\cite{kaufmanandco:scaling,TamaraContainerAnalytisch} and is generalized in the present paper such that it can be used for out-degree distributions that are not Poissonian.  This method is
based on the assumption that almost all frozen nodes can be obtained
by starting from the nodes with a constant update function and by
determining iteratively all nodes that become frozen because some of
their inputs are frozen. This assumption is valid in many cases, in particular for networks that have only constant and reversible functions. If the assumption is not correct, one can nevertheless expect the same scaling exponents, but must use other methods to obtain the frozen core \cite{MoellerFormationFrozenCore}. The main idea of the method is to not specify
the network in advance, but to choose the connections within the
network while determining the frozen core. 
We used the algorithm for our analytical calculations as well as for our computer simulations. The steps of the algorithm are as follows:

\begin{enumerate}

 \item Initially, each of the $N$ nodes is assigned an update function, a number of inputs, and a number of outputs according to the rules given above. These three assignments are made independently from each other, in order to avoid correlations between them.
Nodes with a reversible function are placed into ``containers'' $C_i$ according to their number $k_{in}=i$ of inputs.  Nodes with constant functions are put into the container $C_0$. We denote the number of nodes in container $C_{i}$ by $\left| C_i \right|$. Furthermore, we denote the total number of nodes in all containers by $N_\mathrm{f}=\sum_{i=0}^{i_\mathrm{max}} \left| C_i \right|$, and the total number of inputs to nodes that are not in containers  $C_i$ with $i \ge 1$ by   $k_\mathrm{in}^0$. The initial value of $k_\mathrm{in}^0$ is identical to the number of inputs to nodes with constant functions.  All these values will change during the algorithm. In particular,  $N_\mathrm{f}$ will decrease by 1 with each step.

\item Then, the following steps are iterated until $ \left|C_0 \right| = 0$:
\begin{enumerate}
 \item Select one node from container $C_{0}$. Its number of outputs is denoted by  $k_\mathrm{out}^{N_\mathrm{f}}$.
 \item Choose at random $k_\mathrm{out}^{N_\mathrm{f}}$ out of all $\sum_i i \left|C_i \right| + k_\mathrm{in}^0$ inputs to become connected to the outputs of the selected node. 
\item If $m>0$ inputs of a node in a container $i\ge 1$ became connected to the selected node, move this node from its container $C_{i}$ to $C_{i-m}$. 
\item Reduce the number $k_\mathrm{in}^0$ of unconnected inputs to nodes with constant functions by the number of those that became connected to the selected node in $C_{0}$. This ensures that the total number of outputs in all containers always equals the total number of remaining inputs, $\sum_{i\ge 1} i \left|C_i \right| + k_\mathrm{in}^0$.
 \item  Remove the selected node from the system. This implies the replacement  $N_\mathrm{f}:=N_\mathrm{f}-1$. 

\end{enumerate}

\item The final value of $N_\mathrm{f}$ is identical to the number of nodes that do not belong  to the frozen core of the particular network that was constructed by performing this algorithm. The  probability distribution of  $N_\mathrm{f}$ follows from the stochastic process implemented in this algorithm.

\end{enumerate}

In order to finish the construction of the network, the inputs and outputs of the remaining nodes, which constitute the nonfrozen part of the network, should also be connected at random. However, since we are only interested in the number of nonfrozen nodes and not in the attractors of the networks or the structure of relevant components, we omit this step.

\section{Analytical Considerations}
Based on the algorithm outlined in the previous section, the scaling of the number of nonfrozen nodes with network size can be determined analytically. To this purpose, we define the parameter $\epsilon=\frac{N_\mathrm{f}}{N}$, which is the proportion of nodes that have not yet become frozen. Neglecting fluctuations, the number of nodes in containers $i\ge 1$ can be expressed as \cite{drossel2009critical}
\begin{equation}
\left| C_i \right| = \sum_{l=i}^{i_\mathrm{max}} \left|C_l^\mathrm{ini}\right| \epsilon^i \left( 1-\epsilon\right)^{l-i} \left( l \atop i \right)\, ,\label{eq:Container}
\end{equation}
for $i\ge1$, where $\left|C_l^\mathrm{ini}\right|$ is the number of nodes in container $l$ at the end of step 1.~of the algorithm. When $\epsilon$ is small, only a small proportion of all inputs are still present, and most nodes that are in container $i$ were initially in containers with values $l\gg i$. Therefore, expression (\ref{eq:Container})  can be approximated for small $\epsilon$ by
\begin{equation}
\left| C_i \right| \simeq \epsilon^i \int_i^{i_\mathrm{max}} \left|C_l^\mathrm{ini}\right| e^{-\epsilon l}  l^i dl\, .
\label{eq:Containerapprox}
\end{equation}

Due to the  condition Eq.~(\ref{criticality}) for critical networks, the initial number of nodes in container $C_0 $ is
\begin{equation}
\left| C_0^\mathrm{ini} \right| = \sum_{i=1}^{i_\mathrm{max}} (i-1)\left| C_i^\mathrm{ini} \right| \, , 
\end{equation}

which is equivalent to the condition 
\begin{equation}{N_\mathrm{f}} = \sum_{i=0}^{i_\mathrm{max}} i\left| C_i \right|\, , \end{equation}
which holds during the entire algorithm (again neglecting fluctuations), since on average one input in the containers $i \ge 1$ will become connected to the selected node during one step, because each of the $N_{\mathrm{f}} \langle k \rangle$ remaining inputs connects to the $\langle k \rangle$ outputs of the selected node with the same probability. Consequently we have 
\begin{equation}
\left| C_0\right| = \sum_{i=1}^{i_\mathrm{max}} (i-1)\left| C_i\right| \label{balance}\end{equation}
during the entire algorithm, and the contents of all containers will reach  the value zero at the moment when  $\epsilon$ reaches zero. 

For small values of $\epsilon$, the sum in Eq.~(\ref{balance}) is dominated by the $i=2$ term, as can be concluded from Eq.~(\ref{eq:Container}). This means that for small $\epsilon$ almost all  non-frozen nodes are in $C_1$ and $C_2$ and that 
\begin{equation}
\label{eq:size}
\left|C_{0}\right| \simeq \left|C_{2}\right|
\end{equation}
apart from fluctuations.

Due to random fluctuations, $\left| C_0 \right|$ will typically reach the value zero 
while $N_{\mathrm{f}}$ is still larger than zero. This will happen when the standard deviation of $\left| C_0 \right|$, denoted as $\sigma_{\left|C_{0}\right|}$,  becomes comparable to the average value of $\left| C_0 \right|$, which in turn is identical to the average value of $\left| C_2 \right|$, giving the condition for the end of the process
\begin{equation}
\label{eq:ideaEqual}
\left| C_2 \right|  \simeq \sigma_{\left|C_{0}\right|}. 
\end{equation}
In order to obtain from this relation a condition for the scaling of the final value $N_{\mathrm{f}}$ with $N$, we must express both sides in terms of $N$ and $N_{\mathrm{f}}$ (or, equivalently, $\epsilon$).

In the following, we determine the variance $\sigma_{\left|C_{0}\right|}^2$ of nodes in container $C_0$. In every step exactly one node is removed from the system and some of the nodes from $C_i$ with $i>0$ are moved to $C_0$. The total number of nodes in all containers, $N_{\mathrm{f}}+|C_0|$ decreases exactly by 1 for each iteration of the algorithm. Fluctuations of the total number $N_f$ of nodes in containers $C_i$ with $i>0$ are deviations from the mean values given by (\ref{eq:Containerapprox}). These deviations are identical, but with opposite sign, to fluctuations of the total number of nodes in container $C_0$, since the total number of nodes in the containers decreases in a deterministic manner and can therefore not show random fluctuations. Since for small $\epsilon$ the vast majority of remaining nodes are in $C_1$, the fluctuation of the number of nodes in containers $C_i$ with $i>0$ is dominated by those in container $C_1$, leading to
\begin{equation}
 \sigma_{\left|C_{0}\right|} =  \sigma_{N_\mathrm{f}} \simeq   \sigma_{\left|C_{1}\right|}.
\end{equation}
There are two contributions to this variance:
\begin{enumerate}
 \item The number of nodes $N_\mathrm{f}$ remaining in containers $C_i$ with $i\ge 1$ has to be on average the number of remaining outputs divided by their mean degree. This gives the following contribution to the variance 
\begin{equation}
\sigma^2_{N_\mathrm{f}} \simeq \frac{N_\mathrm{f}}{\langle k \rangle} \sigma^2_{k_\mathrm{out}}\propto N_\mathrm{f} \sigma^2_{k_\mathrm{out}}.
\end{equation}
 \item The second contribution comes from the fact that for small $\epsilon$ the number of nonfrozen inputs is Poisson distributed with a variance identical to the mean value, which is proportional to $N_\mathrm{f}$. Since the vast majority of nodes only have one input, this is also the variance of $\left| C_1\right|$, leading to $\sigma^2_{N_\mathrm{f}} \propto N_\mathrm{f}.$ 
\end{enumerate}
For networks with a scale-free out-degree distribution with an exponent between 2 and 3, the first term dominates and is proportional to $N_{\mathrm{f}}$ multiplied by a power of $N$. Otherwise, the first and second term give together  $\sigma^2_{N_\mathrm{f}} \propto N_\mathrm{f}.$ 
We conclude that
\begin{equation}
 \label{eq:sigma}
 \sigma_{\left|C_{0}\right|} \propto \sqrt{N_\mathrm{f}} \sigma_{k_\mathrm{out}}\propto \sqrt{N_\mathrm{f}} N^A
\end{equation}
with an exponent $A$ that depends on the out-degree distribution. If the variance of the out-degree distribution is finite, we have $A=0$. This holds when the out-degree of all nodes is identical or is Poisson distributed, or when it is a power-law distribution with an exponent $\gamma_\mathrm{out}\ge 3$. When the out-degree distribution is a power law with $\gamma_\mathrm{out} \in(2,3)$, we have 
\begin{equation}
 \sigma_{k_\mathrm{out}} \propto \left(k^\mathrm{max}_\mathrm{out}\right)^{3-\gamma_\mathrm{out}}\, ,
\end{equation}
leading with Eq.~(\ref{kmax1}) or Eq.~(\ref{kmax2}) to $A=3-\gamma_\mathrm{out}$ or $A=(3-\gamma_\mathrm{out})/\gamma_\mathrm{out}$.
For $\gamma_\mathrm{out}=3$, all three expressions agree with each other and give $A=0$, as it must be.

In order to complete the calculation, we need an expression for $\left| C_2 \right|$ in Eq.~(\ref{eq:ideaEqual}). This expression is contained in Eq.~(\ref{eq:Containerapprox}) and was already given in \cite{drossel2009critical} for the case of a Poissonian out-degree distribution. When the in-degree distribution has a finite second moment, which is the case for fixed, Poissonian, or power-law in-degree distributions with $\gamma_\mathrm{in}\ge 3$, the integral in Eq.~(\ref{eq:Containerapprox}) converges even when the cutoff is set to infinity and  $\epsilon$ is set to 0, leading to
\begin{equation}
 \left| C_2 \right|  \propto  N \epsilon^2 \, \label{Eq:C2converge}.
\end{equation}
For networks with a scale-free in-degree distribution we can rewrite Eq.~(\ref{eq:Containerapprox}) as 
\begin{equation}
\left| C_i \right| \propto \epsilon^i N  \int_i^{i_\mathrm{max}}  e^{-\epsilon l}  l^{i-\gamma_\mathrm{in}} dl\, .
\label{eq:Containerapproxpower}
\end{equation}
In the case  $\gamma_\mathrm{in}\ge 3$, the integral converges, leading again to Eq.~(\ref{Eq:C2converge}).
In the case $\gamma_\mathrm{in} \in (2,3)$, the second moment diverges, and the value of the integral is determined either by the upper limit of the integral  $k^\mathrm{max}_\mathrm{in}$ or by the inverse exponential decay constant $\frac{1}{\epsilon}$, whichever is smaller. This gives
\begin{equation}
 \left| C_2\right|  \propto  N \epsilon^2 \left( \mathrm{min}\left(k^\mathrm{max}_\mathrm{in}, \frac{1}{\epsilon}\right)\right)^ {3-\gamma_\mathrm{in}}\, . \label{eq:minInCutoff}
\end{equation}
For the case $k^\mathrm{max}_\mathrm{in}>\frac{1}{\epsilon}$, we obtain \begin{equation}
 \left| C_2\right|  \propto  N \epsilon^{\gamma_\mathrm{in} -1 }\, ,
\end{equation}
independently of the scaling of $k^\mathrm{max}_\mathrm{in}$ with $N$. For  $k^\mathrm{max}_\mathrm{in}<\frac{1}{\epsilon}$, we obtain for $k^\mathrm{max}_\mathrm{in}\propto N$
\begin{equation}
 \left| C_2 \right|  \propto N^{4-\gamma_\mathrm{in}} \epsilon^2\, ,
\end{equation}
and for $k^\mathrm{max}_\mathrm{in}\propto N^{1/\gamma_\mathrm{in}}$
\begin{equation}
 \left| C_2 \right|  \propto N^{\frac{3}{\gamma_\mathrm{in}}} \epsilon^2\, .
\end{equation}
For $\gamma_\mathrm{in}=3$, all four expressions for $|C_2|$ give the same result, as it must be. 

Taking the four cases together, we can write 
\begin{equation}
  \label{eq:generalC2}
  \left| C_2 \right|  \propto  N^a \epsilon^b\,
\end{equation}
with different values for $a$ and $b$.

Inserting this result together with  Eq.~(\ref{eq:sigma}) into Eq.~(\ref{eq:ideaEqual}), we obtain the desired scaling law 
of the number of non-frozen nodes $N_\mathrm{nf}$, which is identical to the final value of $N_{\mathrm{f}}$, as
\begin{equation}
 N_\mathrm{nf} \propto \epsilon N \propto N^{1-\frac{\frac{1}{2}+A-a}{\frac{1}{2} -b}}\equiv N^x\, . \label{eq:scalingNF}
\end{equation}

Since both $k^\mathrm{max}_\mathrm{in}$ and $\frac{1}{\epsilon}$ scale with $N$ in a nontrivial way, and since the scaling of $\epsilon$ with $N$ depends on the result of Eq.~(\ref{eq:minInCutoff}), it is not always possible to tell in advance whether  $k^\mathrm{max}_\mathrm{in}$ is larger or smaller than $\frac{1}{\epsilon}$. In such cases, we assumed first one version of the inequality, and if this gave an inconsistency, we used the other version.  If we write $k^\mathrm{max}_\mathrm{in}\propto N^z$
and use the fact that $\epsilon\propto N^{x-1}$ (with $x$ defined in eq.~(\ref{eq:scalingNF})), the consistency condition reads
\begin{equation}
k^\mathrm{max}_\mathrm{in}\lessgtr\frac{1}{\epsilon}  \Leftrightarrow z+x-1\lessgtr0 \, .\label{eq:maxCheck}
\end{equation}

\begin{figure}
\includegraphics[width=1\linewidth,page=1]{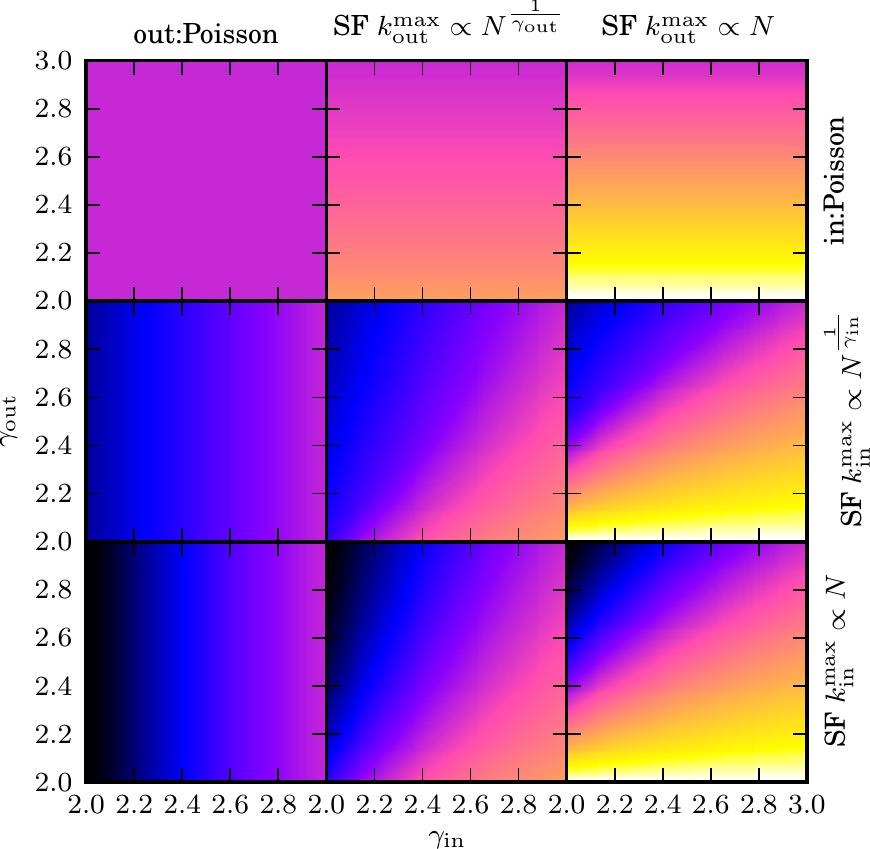}

\vspace{2mm}

\includegraphics[width=1\linewidth,page=2]{exponentExamples3d.pdf}

\vspace{2mm}

\includegraphics[width=1\linewidth,page=4]{exponentExamples3d.pdf}
\caption{(Color online) Graphical representation of the nine cases listed in Table 1. Top: exponent $x$. Bottom: exponent $\frac{y}{x}$.}
\label{fig:exponentTheory}
\end{figure}

In the standard case, where the second moments of the in- and out-degree distributions are finite, we have $A=0$, $a=1$, and $b=2$, leading to the well-known result
\begin{equation}
 N_\mathrm{nf} \propto N^\frac{2}{3}\, .
\end{equation}
Since we consider three cases for both the in-degree and the out-degree distribution, there are altogether 9 different relations for the scaling of the number of non-frozen nodes with the total number of nodes in a critical network. These 9 cases are listed in Tab.~\ref{tab:exponents} and represented graphically in Fig.~\ref{fig:exponentTheory}. The most striking feature of these results is that the scaling exponent for the number of nonfrozen nodes increases with $\gamma_\mathrm{in}$, but decreases with $\gamma_\mathrm{out}$. In the case of a scale-free out-degree disribution, the exponent characterizing the proportion of nonfrozen nodes that have two nonfrozen inputs increases with decreasing $\gamma_\mathrm{out}$.

There are several ways to check the correctness of these expressions:
\begin{itemize}
 \item Every expression for $x$ must give $\frac{2}{3}$ for $\gamma_\mathrm{in} = \gamma_\mathrm{out}=3$. If both distributions are scale free, $x$ must take the expression for the corresponding Poisson distribution when one of the $\gamma$ values is set to 3. 
 \item $x$ must be in the interval $\left[0,1\right]$ for $\gamma\in\left[2,3\right]$.
 \item For a scale-free in-degree distribution, one of the two possible expressions for $x$ should fulfill the consistency condition Eq.~(\ref{eq:maxCheck}). If both fulfill the condition, the values of $x$ must be identical in the two cases.
 \item From Eq.~(\ref{eq:maxCheck}) follows that  $x=1-z$ fo the border case. For $z=1$ ($k^\mathrm{max}_\mathrm{in}\propto N$), this meas that $x=0$ is only possible on the border of the considered range. For $z=\frac{1}{\gamma_\mathrm{in}}$, the values of $x$ must be  $x=1-\frac{1}{\gamma_\mathrm{in}}$ on the border, independently of the out-degree distribution.
\end{itemize} 
We also did the ultimate check, which is computer simulations, as described in section \ref{sec:experiments}.

In a similar way, one obtains scaling laws for $\left|\mathrm{C}_2\right|$. To this purpose, the scaling of $\epsilon$ with $N$ from Eq.~(\ref{eq:scalingNF}) must be combined with Eq.~(\ref{eq:generalC2}), giving
\begin{equation}
 \left|\mathrm{C}_2\right| \propto N^{a+b\frac{\frac{1}{2}+A-a}{b-\frac{1}{2}}}\equiv N^y. \label{eq:scalingC2}
\end{equation}
The values of $y$ for the nine cases are also listed in Tab.~\ref{tab:exponents}, and $\frac{y}{x}$ is visualized in Fig.~\ref{fig:exponentTheory}. The check for correctness can be made in the same way as for the exponent $x$, with the only difference that $y$ must be $\frac{1}{3}$ for $\gamma = 3$.

\begin{table*}[ht]
% \vspace{15cm}
% \hfill
\begin{tabular}{|l|c|c||c|c|c|c|c|}
\hline 

$x$ & \multicolumn{2}{c||}{out $\rightarrow$} & Poisson & \multicolumn{2}{c|}{SF $k^\mathrm{max}_\mathrm{out}\propto N^{\frac{1}{\gamma_\mathrm{out}}}$} &  \multicolumn{2}{c|}{SF $k^\mathrm{max}_\mathrm{out}\propto N$} \tabularnewline
\hline 
in $\downarrow$ & \multicolumn{2}{c||}{} & $\sigma_{\left|k_\mathrm{out}\right|}\propto1$ & \multicolumn{2}{c|}{$\sigma_{\left|k_\mathrm{out}\right|}\propto N^{\frac{3-\gamma_\mathrm{out}}{2\gamma_\mathrm{out}}}$} & \multicolumn{2}{c|}{$\sigma_{\left|k_\mathrm{out}\right|}\propto N^\frac{3-\gamma_\mathrm{out}}{2}$} \tabularnewline
\hline 
\hline 
Poisson & \multicolumn{2}{c||}{$\left|C_{2}\right|\propto N\epsilon^{2}$} & $\frac{2}{3}$ & \multicolumn{2}{c|}{$\frac{1}{3}+\frac{1}{\gamma_\mathrm{out}}$} & \multicolumn{2}{c|}{$\frac{5-\gamma_\mathrm{out}}{3}$} \tabularnewline
\hline 
\multirow{3}{*}{SF $k^\mathrm{max}_\mathrm{in}\propto N^{\frac{1}{\gamma_\mathrm{in}}}$} &
$\left|C_{2}\right|\propto N^{\frac{3}{\gamma_\mathrm{in}}}\epsilon^{2}$ & $x\le1-\frac{1}{\gamma_\mathrm{in}}$ &
\multirow{3}{*}{$\frac{4}{3}-\frac{2}{\gamma_\mathrm{in}}$} &
          $ 1 + \frac{1}{\gamma_\mathrm{out}}-\frac{2}{\gamma_\mathrm{in}}$ & $\gamma_\mathrm{in}\le\gamma_\mathrm{out}$ &
       $\frac{7-\gamma_\mathrm{out}}{3}-\frac{2}{\gamma_\mathrm{in}}$ & $\gamma_\mathrm{in}\le\frac{3}{4-\gamma_\mathrm{out}}$  \tabularnewline
\cline{5-8}  &  &  & &
          $ 1-\frac{1}{\gamma} $ & $\gamma=\gamma_\mathrm{in}=\gamma_\mathrm{out}$ &
          $1-\frac{1}{\gamma_\mathrm{in}}$ & $\gamma_\mathrm{in}=\frac{3}{4-\gamma_\mathrm{out}}$ 
       \tabularnewline
\cline{5-8}  & $\left|C_{2}\right|\propto N\epsilon^{\gamma_\mathrm{in}-1}$  & $x\ge1-\frac{1}{\gamma_\mathrm{in}}$ & &
       $1-\frac{1-\frac{3-\gamma_\mathrm{out}}{\gamma_\mathrm{out}}}{2\gamma_\mathrm{in}-3}$& $\gamma_\mathrm{in}\ge\gamma_\mathrm{out}$&
          $1-\frac{2-\gamma_\mathrm{out}}{3-2\gamma_\mathrm{in}}$ & $\gamma_\mathrm{in}\ge\frac{3}{4-\gamma_\mathrm{out}}$  \tabularnewline
\hline 
\multirow{2}{*}{SF $k^\mathrm{max}_\mathrm{in}\propto N$} & 
\multicolumn{2}{c||}{\multirow{2}{*}{$\left|C_{2}\right|\propto N\epsilon^{\gamma_\mathrm{in}-1}$}} &
\multirow{2}{*}{$\frac{2\gamma_\mathrm{in}-4}{2\gamma_\mathrm{in}-3}$} &
       $1-\frac{1-\frac{3-\gamma_\mathrm{out}}{\gamma_\mathrm{out}}}{2\gamma_\mathrm{in}-3}$& always &
          $1-\frac{2-\gamma_\mathrm{out}}{3-2\gamma_\mathrm{in}}$ & always  \tabularnewline
\cline{5-8} &\multicolumn{2}{c||}{}  & &
          $ 1-\frac{1}{\gamma} $ & $\gamma=\gamma_\mathrm{in}=\gamma_\mathrm{out}$ &
          $\frac{\gamma-1}{2\gamma-3}$ & $\gamma=\gamma_\mathrm{in}=\gamma_\mathrm{out}$   \tabularnewline
\hline 

\end{tabular}
\hfill \hfill

\vspace{0.2cm}

% \hfill
% \hspace*{-1.3cm}
\begin{tabular}{|l|c|c||c|c|c|c|c|}
\hline 
$y$ & \multicolumn{2}{c||}{out $\rightarrow$} & Poisson & \multicolumn{2}{c|}{SF $k^\mathrm{max}_\mathrm{out}\propto N^{\frac{1}{\gamma_\mathrm{out}}}$} &  \multicolumn{2}{c|}{SF $k^\mathrm{max}_\mathrm{out}\propto N$} \tabularnewline
\hline 
in $\downarrow$ & \multicolumn{2}{c||}{} & $\sigma_{\left|k_\mathrm{out}\right|}\propto1$ & \multicolumn{2}{c|}{$\sigma_{\left|k_\mathrm{out}\right|}\propto N^{\frac{3-\gamma_\mathrm{out}}{2\gamma_\mathrm{out}}}$} & \multicolumn{2}{c|}{$\sigma_{\left|k_\mathrm{out}\right|}\propto N^\frac{3-\gamma_\mathrm{out}}{2}$} \tabularnewline
\hline 
\hline 
Poisson & \multicolumn{2}{c||}{$\left|C_{2}\right|\propto N\epsilon^{2}$} & $\frac{1}{3}$ & \multicolumn{2}{c|}{$\frac{2}{\gamma_\mathrm{out}}-\frac{1}{3}$} & \multicolumn{2}{c|}{$\frac{7-2\gamma_\mathrm{out}}{3}$} \tabularnewline
\hline 
\multirow{3}{*}{SF $k^\mathrm{max}_\mathrm{in}\propto N^{\frac{1}{\gamma_\mathrm{in}}}$} &
$\left|C_{2}\right|\propto N^{\frac{3}{\gamma_\mathrm{in}}}\epsilon^{2}$ & $x\le1-\frac{1}{\gamma_\mathrm{in}}$ &
\multirow{3}{*}{$\frac{2}{3}-\frac{1}{\gamma_\mathrm{in}}$} &
          $ \frac{2}{\gamma_\mathrm{out}}-\frac{1}{\gamma_\mathrm{in}} $  & $\gamma_\mathrm{in}\le\gamma_\mathrm{out}$ &
       $\frac{8-2\gamma_\mathrm{out}}{3}-\frac{1}{\gamma_\mathrm{in}}$ & $\gamma_\mathrm{in}\le\frac{3}{4-\gamma_\mathrm{out}}$ \tabularnewline
\cline{5-8}  &  &  & &
          $ \frac{1}{\gamma}$  &$\gamma=\gamma_\mathrm{in}=\gamma_\mathrm{out}$ &
          $\frac{1}{\gamma_\mathrm{in}}$ &$\gamma_\mathrm{in}=\frac{3}{4-\gamma_\mathrm{out}}$ 
       \tabularnewline
\cline{5-8}  & $\left|C_{2}\right|\propto N\epsilon^{\gamma_\mathrm{in}-1}$  & $x\ge1-\frac{1}{\gamma_\mathrm{in}}$ & & 
       $1-\frac{\left(\gamma_\mathrm{in}-1\right)\left(2-\gamma_\mathrm{out}\right)}{\gamma_\mathrm{out}\left(3-2\gamma_\mathrm{in}\right)}$ & $\gamma_\mathrm{in}\ge\gamma_\mathrm{out}$ &
          $1-\frac{\left(\gamma_\mathrm{in}-1\right)\left(2-\gamma_\mathrm{out}\right)}{3-2\gamma_\mathrm{in}}$& $\gamma_\mathrm{in}\ge\frac{3}{4-\gamma_\mathrm{out}}$   \tabularnewline
\hline 
\multirow{2}{*}{SF $k^\mathrm{max}_\mathrm{in}\propto N$} & 
\multicolumn{2}{c||}{\multirow{2}{*}{ $\left|C_{2}\right|\propto N\epsilon^{\gamma_\mathrm{in}-1}$ }} &
\multirow{2}{*}{$\frac{\gamma_\mathrm{in}-2}{2\gamma_\mathrm{in}-3}$} & 
       $1-\frac{\left(\gamma_\mathrm{in}-1\right)\left(2-\gamma_\mathrm{out}\right)}{\gamma_\mathrm{out}\left(3-2\gamma_\mathrm{in}\right)}$ & always&
          $1-\frac{\left(\gamma_\mathrm{in}-1\right)\left(2-\gamma_\mathrm{out}\right)}{3-2\gamma_\mathrm{in}}$& always   \tabularnewline
\cline{5-8}  & \multicolumn{2}{c||}{}  & &
          $\frac{1}{\gamma} $ &$\gamma=\gamma_\mathrm{in}=\gamma_\mathrm{out}$ &
          $\frac{5-5\gamma+\gamma^2}{3-2\gamma}$ &$\gamma=\gamma_\mathrm{in}=\gamma_\mathrm{out}$ \tabularnewline
\hline 
\end{tabular}
\hfill %\hfill

\caption{ \label{tab:exponents} The 9 different expressions for the exponents $x$ and $y$ that characterize the scaling $N_\mathrm{nf}\propto N^{x}$ of the number of nonfrozen nodes $N_\mathrm{nf}$ with $N$ and the scaling of the number of nonfrozen nodes with two nonfrozen inputs $\left|\mathrm{C}_2\right|\propto N^{y}$ with $N$. The expressions obtained in the special case $\gamma_\mathrm{in}=\gamma_\mathrm{out}$ are also specified. Where necessary, two expressions for $x$ and the corresponding conditions are given, as well as the boundary case for which both expressions hold simultaneously. }

\end{table*}

\section{Extension to Other Sets of Update Functions}

All results so far were derived by using only constant and reversible update functions. However, the algorithm can be generalized to more general sets of update functions, in particular to biased functions. For general sets of functions, in step 1.~nodes with constant functions are placed in container $C_0$, while all other nodes are placed in containers according to their number of inputs. Also step 2.~has to be modified. When $m \ge 1$ inputs of a node in container $i$ become connected to the selected node, the node in container $i$ may freeze completely and is then moved to container $C_0$ instead of $C_{i-m}$. This occurs with a probability that depends on the set of update functions and is given by the probability that a randomly chosen function of $i$ inputs becomes constant when $m$ of the inputs become frozen. In this case also the value of $k_\mathrm{in}^0$ has to be increased by $i-m$. Clearly, the container method only works when the probability distribution of the update functions with $i$ 
inputs is identical to the conditional probability distribution that is obtained by freezing $l-i$ inputs of nodes that are initially in a container $C_l$ with  $l>i$. 

The analytical considerations become slightly different, but lead to the same conclusions.  Eq.~(\ref{eq:Container}) still states that towards the end of the algorithm only containers  $C_0$, $C_1$, and $C_2$ need to be considered.  Nodes in $C_0$ or $C_1$ will behave in the same way as for the case of constant and reversible update functions. The only difference is that a certain proportion of the nodes in $C_2$ become already frozen if only one of their inputs is connected to a frozen node. Such nodes could in principle be placed in container $C_1$, and then the situation would be identical to the one of only constant and reversible functions. Due to the fact that $|C_1|\gg |C_2|$, this make no significant difference to the calculations  and therefore has no effect to the scaling laws.

\section{Computer simulations}
\label{sec:experiments}
\begin{figure}[t!]
 \includegraphics[trim=0.07cm 0.10cm 0.75cm 0.80cm, clip=true,page=12]{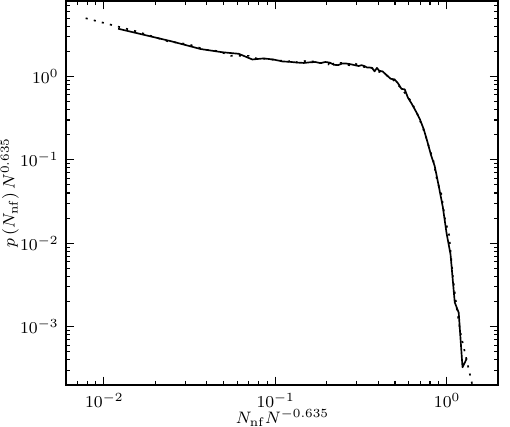}
\caption{Typical example for the probability distribution of nonfrozen nodes.
This sample is for critical networks with a scale free in- and out-degree distribution with $\gamma=2.3$ and a cutoff scaling as $k^\mathrm{max}\propto N^\frac{1}{\gamma}$. The number of realisations was $2\cdot10^5$, and number of nodes in the network $N=2^{10}$ for the solid and $N=2^{11}$ for the dotted curve. The curves  were collapsed using the exponent $0.635$, which is the best value our algorithm found for these two curves.}
\label{fig:typResult}
\end{figure}

In order to confirm our analytical calculations, we performed  computer simulations of the algorithm. Instead of a Poisson distribution we used a fixed value for $k$. If the value of $\left<k\right>$ required for criticality was not an integer, we used a mixture of the two neighbouring integer values for generating the distribution of $k$ values.
Avoiding unwanted correlations between in-degree and out-degree distribution or between a degree distribution and the distribution of Boolean functions in the case of scale-free distributions turned out to be challenging. The 
simulations were done in the following way. For each set of in- and out-degree distributions and each value of $\gamma$, we run the algorithm between $10^3$ and $10^6$ times (the smaller value applies to larger values of $N$, due computation costs) for values of $N\in\left\{2^{10},2^{11},\ldots,2^{20},\ldots\right\}$ and thus obtained a distribution for the number of nonfrozen $N_\mathrm{nf}$ and the final size of nodes in container $C_2$. The data were smoothened by performing a logarithmical binning with a step width of $1.05$. To further smoothen the first part of the curve where the slope is very small (in the log-log plot), we combined groups of neighboring bins that comprise approximately the same number of events. A typical result is shown in Fig.~\ref{fig:typResult}, where the axes were scaled using the exponent that gives the best data collapse.

\begin{figure}[t!]
 \includegraphics[trim=1.15cm 0.90cm 0.90cm 0.60cm, clip=true,page=2]{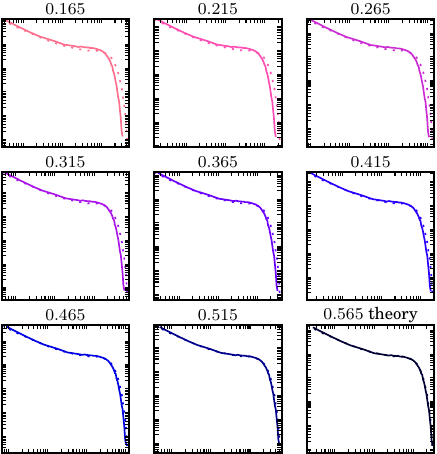}
\caption{(Color online) Example for the color coding in Fig.~\ref{fig:pcolor}. Each subplot is similar to Fig.~\ref{fig:typResult}. The only difference is the number of nodes, which was in this case $2^{17}$ and $2^{18}$. The numbers above the graphs give the exponents used for scaling the two curves, and the color indicates the quality of the collapse. The theoretical exponent is used in the last plot.}
\label{fig:colSamples}
\end{figure}

However, the quality of the data collapse is not always that good, in particular when $\gamma$ is close to 2 or when $N$ is small. 
In order to compare the analytical expression, which should become exact in the limit $N\to \infty$, to the simulation results, we established an automated procedure for quantifying the quality of the date collapse between two curves for system sizes $N_1=2^n$ and $N_2=2^{n+1}$ for the entire possible interval $(0,1)$. This automated procedure gives an optimal ($N$-dependent) value of the exponent, which should approach the theoretical value as $N$ increases. The quality of the collapse was quantified by a fitness value, which is the mean of the absolute value of the distance between the two linearly interpolated curves, omitting the first 20 percent of the curves. We convinced ourselves that higher fitness values correspond to what is intuitively considered a better collapse. The results are shown in Fig.~\ref{fig:pcolor}, using colors to indicate the fitness. Each box shows on the $y$ axis the scaling exponent used for the collapse and on the $x$ axis the value of $N_1$ of the pair that was compared. The 
color 
scale was set such that the interval between the minimum and maximum fitness value was stretched to $[0,1]$
and divided by 0.7 and then taken to the  power of 0.5 to improve color resolution for minimal (optimal) fitness values close to 0. Fig.~\ref{fig:colSamples} gives an impression of the quality of the collapse associated with the different colors. Some of the boxes contain white areas due to missing simulation values or to  large fluctuations due to poor statistics. The theoretical value for the exponent lies in almost all boxes in the  area of best fitness values. In some of the cases one can see that this area is moving down or up with increasing system size $N_1$. This indicates finite size effects. 

We also did this evaluation for the scaling exponents for $\left|C_2\right|$ (not shown), and we found again that the agreement between theory and computer simulations is very good.

\newlength{\firstSp}
\setlength{\firstSp}{2.8cm}

\newlength{\secSp}
\setlength{\secSp}{1.4cm}

\newlength{\wid}
\setlength{\wid}{3.6cm}

\newlength{\imwid}
\setlength{\imwid}{0.52\textwidth}

\begin{figure*}[t!]
\begin{minipage}[b]{\wid}
in: SF $k^\mathrm{max}_\mathrm{in}\propto N^\frac{1}{\gamma_\mathrm{in}}$\\
out: Fixed degree
\vspace*{\firstSp}\\
in: SF $k^\mathrm{max}_\mathrm{in}\propto N$\\
out: Fixed degree
\vspace*{\secSp}
\end{minipage}\includegraphics[trim=2cm 1.5cm 2cm 1.0cm, clip=true,width=\imwid]{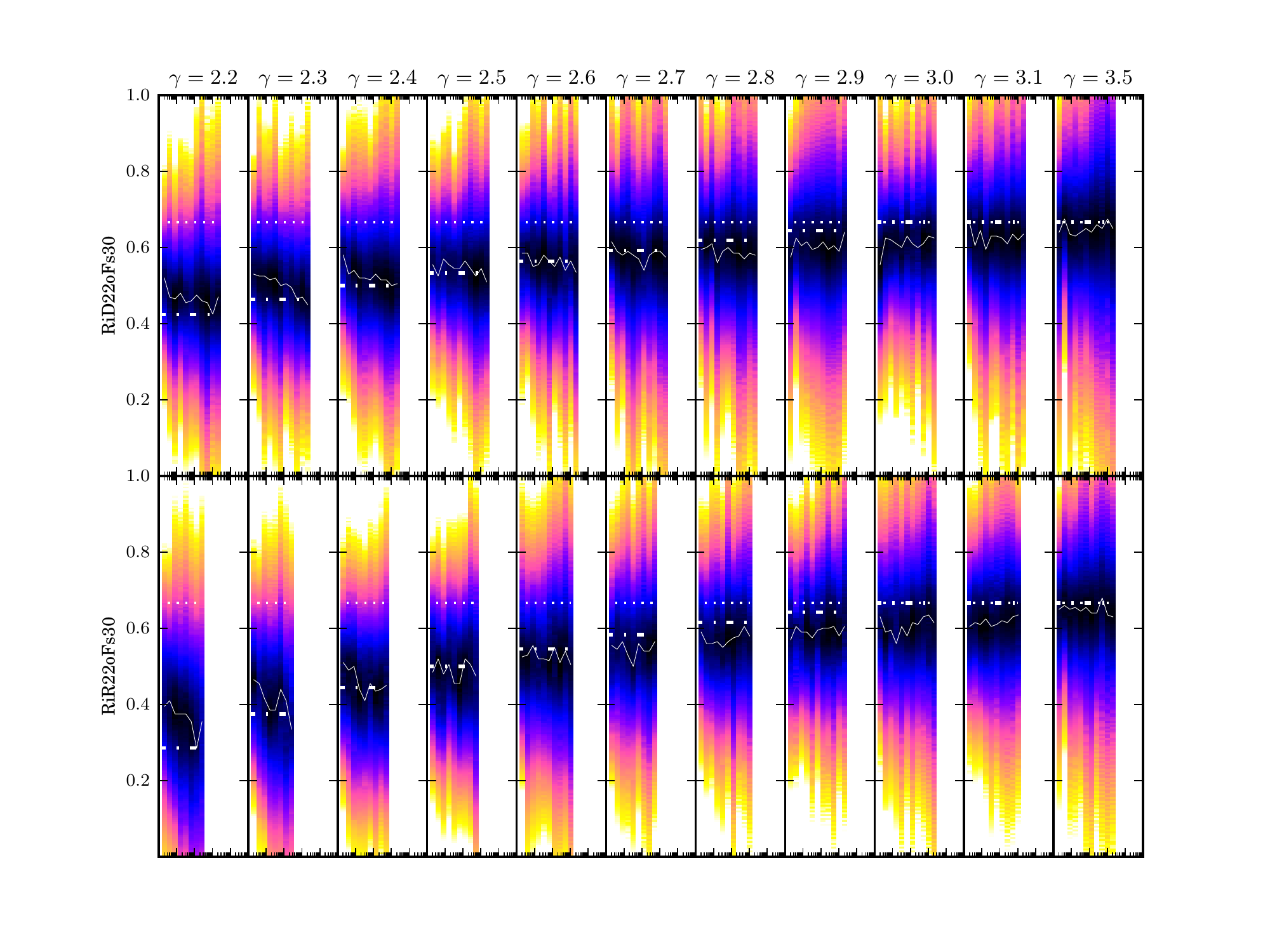}\newline
\begin{minipage}[b]{\wid}
in: Fixed degree\\
out: SF $k^\mathrm{max}_\mathrm{out}\propto N^\frac{1}{\gamma_\mathrm{out}}$
\vspace*{\firstSp}\\
in: Fixed degree\\
out: SF $k^\mathrm{max}_\mathrm{out}\propto N$
\vspace*{\secSp}
\end{minipage}\includegraphics[trim=2cm 1.5cm 2cm 1.4cm, clip=true,width=\imwid]{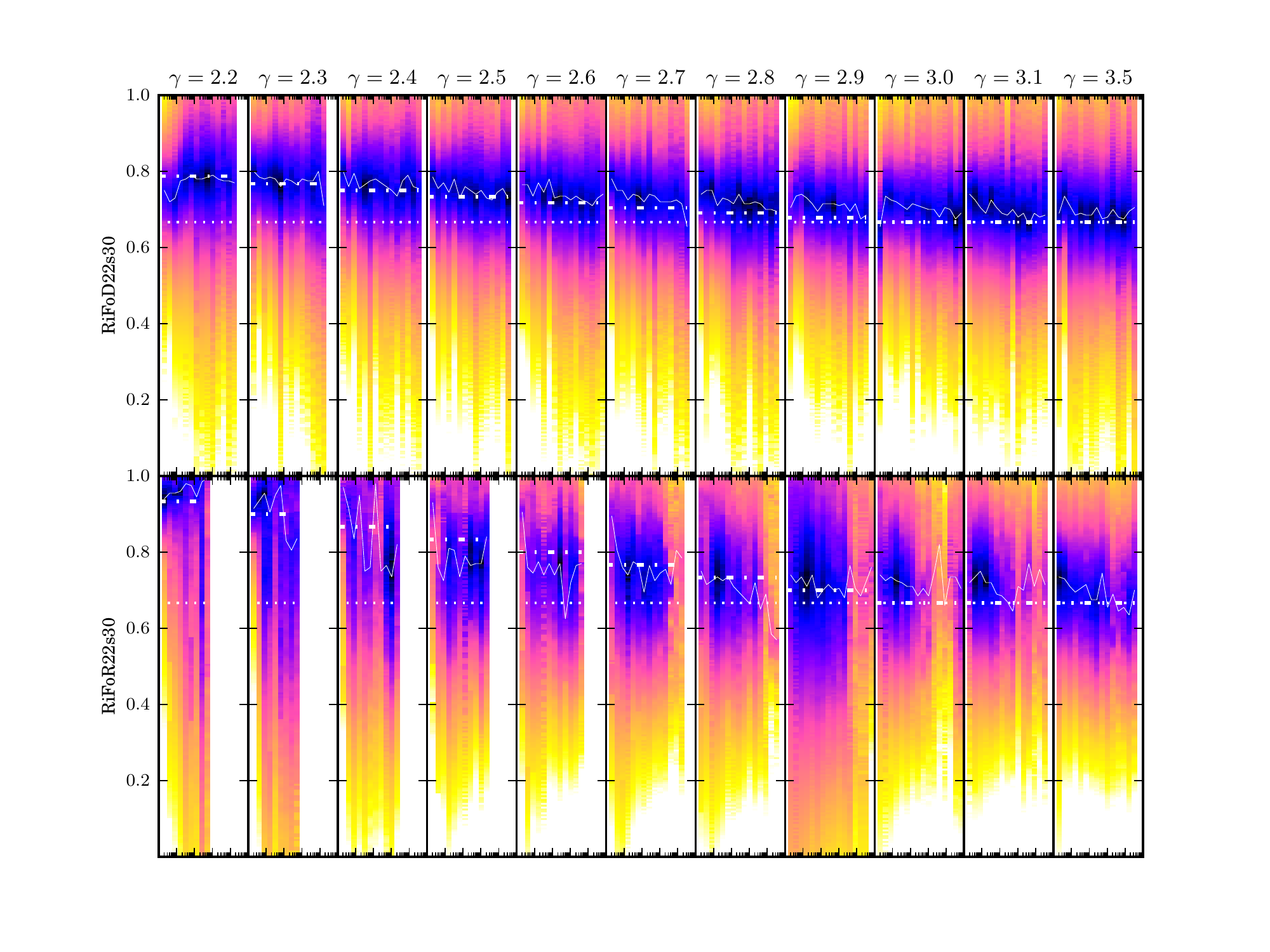}\newline
\begin{minipage}[b]{\wid}
in: SF $k^\mathrm{max}\propto N^\frac{1}{\gamma}$\\
out: SF $k^\mathrm{max}\propto N^\frac{1}{\gamma}$
\vspace*{\firstSp}\\
in: SF $k^\mathrm{max}\propto N$\\
out: SF $k^\mathrm{max}\propto N$
\vspace*{\secSp}
\end{minipage}\includegraphics[trim=2cm 1.5cm 2cm 1.4cm, clip=true,width=\imwid]{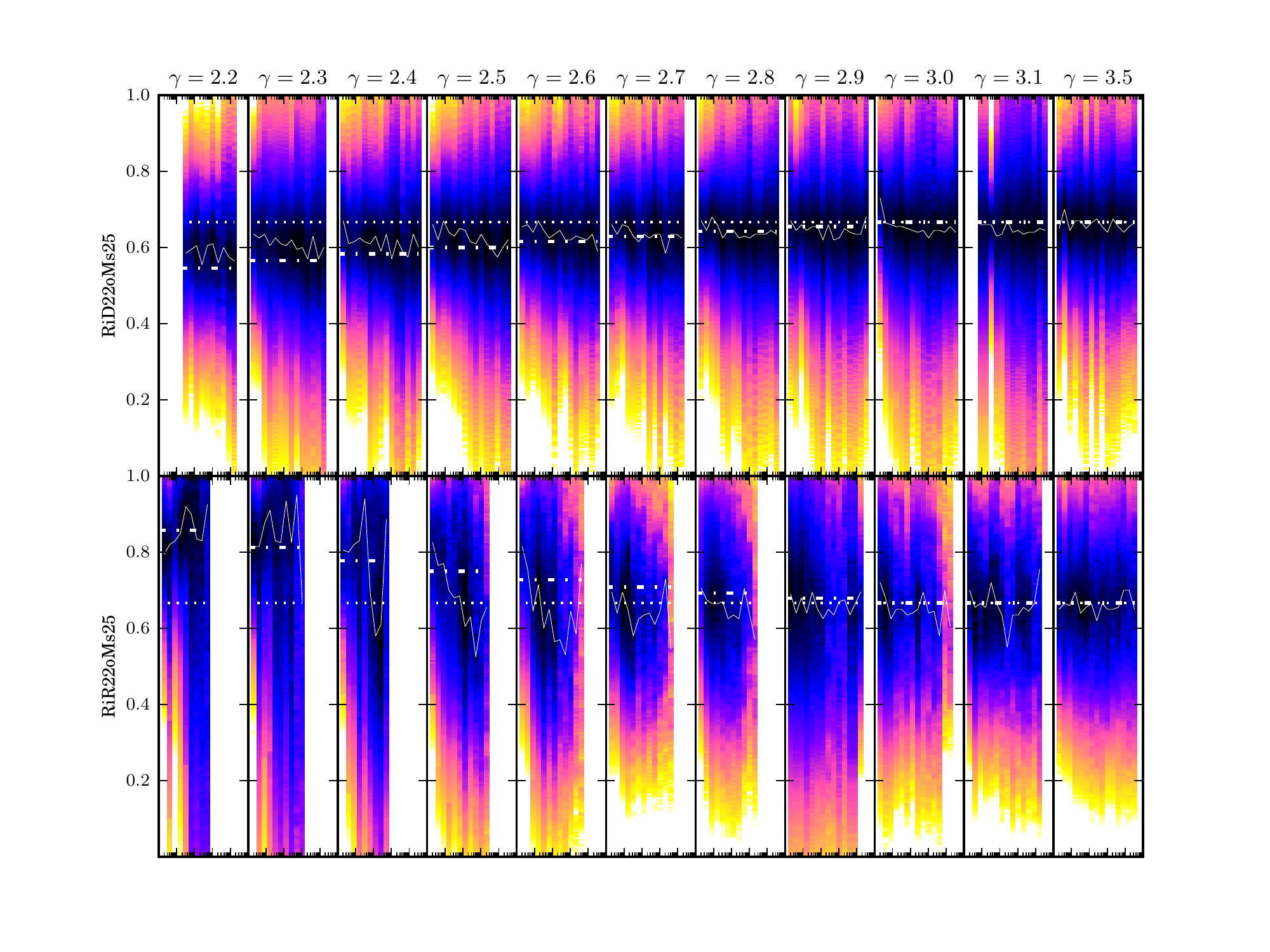}\newline
\vspace*{-0.5cm}
\caption{(Color online) The quality of the data collapse (color coded)  of the distributions $P(N_{\mathrm nf})$ obtained for two $N$ values that differ by a factor of 2, as a function of $N$ ($x$-axis log.  $10^3$ to $10^8$) and the exponent used for the collapse ($y$ axis lin. $0$ to $1$). The darker the color, the better the collapse. The different plots are for different combinations of the in- and out-degree distributions. When the in-degree distribution was identical to the out-degree distribution, we generated the out-degree distribution by using exactly the same sample as the one obtained for the in-degree distribution.  The lines give the value 2/3 (dotted line), the theoretically predicted exponent (dash-dotted line), and the best exponent for the collapse of the pairs of curves (solid line).}\label{fig:pcolor}
\end{figure*}

\section{Conclusions}

Using analytical calculations and computer simulations, we have determined the scaling of the number of nonfrozen nodes of critical scale-free random Boolean networks with network size, leading to a stunning variety of different scaling laws. Our calculations are based on an algorithm that determines the frozen core of the network by an iterative procedure. The scaling exponent for the number of nonfrozen nodes does not only depend on the values of two exponents of the power-law degree distributions (if they are smaller than 3), but also on whether these scale-free distributions are generated by randomly sampling the degree of each node from a power-law distribution or by exactly matching the degree distribution of each realization of the network to the desired power law. We calculated the scaling laws by generalizing a phenomenological theory that has been used earlier for critical random Boolean networks with Poissonian out-degree distributions. Furthermore, we performed computer simulations using a very 
efficient algorithm that allowed us to test the theoretical results for network sizes larger than $2^{20}$. These computer simulations confirm our analytical considerations.
This work  thus fills an important gap in our understanding of scale-free critical random Boolean network. 

Our results show that the size of the nonfrozen part of the critical network decreases with decreasing $\gamma_\mathrm{in}$ but increases with decreasing $\gamma_\mathrm{out}$. In the case where the cutoff of the out-degree distribution scales as $N$, the scaling exponent of the nonfrozen part of the network approaches 1 as $\gamma_\mathrm{out}$ decreases towards 2. This means that a finite proportion of all nodes remain nonfrozen in this limit case. In the case $\gamma_\mathrm{in} =\gamma_\mathrm{out}\equiv \gamma$, the trend of the scaling exponents with $\gamma$ depends on the scaling of the cutoffs with network size.  The opposite trends observed for scale-free in- and out-degree distributions can be explained as follows: a smaller value of $\gamma_\mathrm{in}$ leads to smaller fluctuations in the contents of the containers (i.e., in the growth of the frozen core) and therefore to a later stop of the freezing algorithm, while a smaller value of $\gamma_\mathrm{out}$ leads to larger fluctuations in the 
contents of the containers and therefore to an earlier stop of the freezing algorithm. 

The proportion of nonfrozen nodes with two nonfrozen inputs increases 
as $\gamma_\mathrm{out}$ decreases towards 2, and it approaches the value 1 in the case where the cutoff of the out-degree distribution is proportional to $N$.
In contrast, in the case of Poissonian out-degree distribution the number of nonfrozen nodes with two nonfrozen inputs scales as the square root of the total number of nonfrozen nodes, irrespective of the in-degree distribution. For this case, it was shown in \cite{kaufman2006relevant} that the computational core of the network (i.e., the set of \emph{relevant} nodes), which determines the number and length of attractors, scales also as the square root of the number of nonfrozen nodes. Our results show that the computational core increases when the out-degree distribution becomes scale free, and the majority of relevant components are not any more simple loops. This means that the length of attractors is much larger for scale-free out-degree distributions than for scale-free in-degree distributions. 

Finally, we want to emphasize that we only investigated situations where the in- and out-degree of a nodes were uncorrelated. However, the case where the two degrees are correlated (they can for instance be identical) is also relevant. It applies in particular to networks that have undirected connections, which means that the in- and outgoing connections are identical. As argued in \cite{fronczak.fronczak.ea:kauffman}, the undirected case with a degree-distribution exponent $\gamma$ corresponds to the uncorrelated directed case with an exponent $\gamma-1$.

\section*{acknowledgments}

This work was supported by the DFG under grant number Dr300/4-2.

\bibliographystyle{unsrt}
\bibliography{literatur}

\begin{thebibliography}{10}

\bibitem{kauffman:random}
Stuart Kauffman, Carsten Peterson, Bj\o{}rn Samuelsson, and Carl Troein.
\newblock Random {B}oolean network models and the yeast transcriptional
  network.
\newblock {\em Proc. Natl. Acad. Sci. U.S.A.}, 100:14796, 2003.

\bibitem{kauffman:metabolic}
Stuart~A. Kauffman.
\newblock Metabolic stability and epigenesis in randomly constructed genetic
  nets.
\newblock {\em J. Theor. Biol.}, 22:437--467, January to March 1969.

\bibitem{BornholdtScience}
S.~Bornholdt.
\newblock Less is more in modeling large genetic networks.
\newblock {\em Science}, 310(5747):449, 2005.

\bibitem{reviewbarbara}
B.~Drossel.
\newblock Random boolean networks.
\newblock {\em Reviews of nonlinear dynamics and complexity}, pages 69--110,
  2008.

\bibitem{kauffman:homeostasis}
Stuart Kauffman.
\newblock Homeostasis and differentation in random genetic control networks.
\newblock {\em Nature}, 224:177--178, October 11 1969.

\bibitem{Kauffman1969437}
S.A. Kauffman.
\newblock Metabolic stability and epigenesis in randomly constructed genetic
  nets.
\newblock {\em Journal of Theoretical Biology}, 22(3):437 -- 467, 1969.

\bibitem{socolar:scaling}
Joshua E.~S. Socolar and Stuart~A. Kauffman.
\newblock Scaling in ordered and critical random {B}oolean networks.
\newblock {\em Phys. Rev. Lett.}, 90:068702, 2003.

\bibitem{kaufmanandco:scaling}
Viktor Kaufman, Tamara Mihaljev, and Barbara Drossel.
\newblock Scaling in critical random boolean networks.
\newblock {\em Physical Review E}, 72(4):046124, 2005.

\bibitem{TamaraContainerAnalytisch}
Tamara Mihaljev and Barbara Drossel.
\newblock Scaling in a general class of critical random boolean networks.
\newblock {\em Phys. Rev. E}, 74(4):046101, Oct 2006.

\bibitem{albert2005scale}
R.~Albert.
\newblock Scale-free networks in cell biology.
\newblock {\em Journal of cell science}, 118(21):4947--4957, 2005.

\bibitem{fox.hill:from}
Jeffrey~J. Fox and Colin~C. Hill.
\newblock From topology to dynamics in biochemical networks.
\newblock {\em Chaos}, 11:809--815, December 2001.

\bibitem{kinoshita.iguchi.ea:robustness}
Shu-ichi Kinoshita, Kazumoto Iguchi, Hiroaki Yamada, Yamada Tokuyama, Michio
  Tokuyama, Irwin Oppenheim, and Hideya Nishiyama.
\newblock Robustness of attractor states in complex networks.
\newblock {\em AIP Conference Proceedings}, 982:768--771, February 2008.

\bibitem{aldana03}
Maximino Aldana.
\newblock Boolean dynamics of networks with scale-free topology.
\newblock {\em Physica D}, 185:45, 2003.

\bibitem{oikonomou2006effects}
P.~Oikonomou and P.~Cluzel.
\newblock Effects of topology on network evolution.
\newblock {\em Nature Physics}, 2(8):532--536, 2006.

\bibitem{2003PNAS..100.8710A}
M.~{Aldana} and P.~{Cluzel}.
\newblock {A natural class of robust networks}.
\newblock {\em Proceedings of the National Academy of Science}, 100:8710--8714,
  July 2003.

\bibitem{fronczak.fronczak.ea:kauffman}
Piotr Fronczak, Agata Fronczak, and Janusz~A. Holyst.
\newblock Kauffman boolean model in undirected scale-free networks.
\newblock {\em Phys. Rev. E}, 77:036119--5, March 2008.

\bibitem{iguchi2005boolean}
K.~Iguchi, S.~Kinoshita, and H.S. Yamada.
\newblock Boolean dynamics of kauffman models with a scale-free network.
\newblock {\em arXiv preprint cond-mat/0510430}, 2005.

\bibitem{drossel2009critical}
B.~Drossel and F.~Greil.
\newblock Critical boolean networks with scale-free in-degree distribution.
\newblock {\em Physical Review E}, 80(2):026102, 2009.

\bibitem{2004PhyA..339..665S}
R.~{Serra}, M.~{Villani}, and L.~{Agostini}.
\newblock {On the dynamics of random Boolean networks with scale-free outgoing
  connections}.
\newblock {\em Physica A Statistical Mechanics and its Applications},
  339:665--673, August 2004.

\bibitem{lee2008broad}
D.S. Lee and H.~Rieger.
\newblock Broad edge of chaos in strongly heterogeneous boolean networks.
\newblock {\em Journal of Physics A: Mathematical and Theoretical}, 41:415001,
  2008.

\bibitem{MoellerFormationFrozenCore}
M.~M\"oller and B.~Drossel.
\newblock The formation of the frozen core in critical boolean networks.
\newblock {\em New Journal of Physics}, 14(2):023051, 2012.

\bibitem{kaufman2006relevant}
V.~Kaufman and B.~Drossel.
\newblock Relevant components in critical random boolean networks.
\newblock {\em New Journal of Physics}, 8(10):228, 2006.

\end{thebibliography}

\end{document}